\documentstyle[aps,multicol,prb,epsf]{revtex}
\begin{document}
\draft
\title{Quantized adiabatic charge pumping and resonant transmission}
\author{O. Entin-Wohlman and Amnon Aharony}
\address{School of Physics and Astronomy, Raymond and Beverly Sackler
Faculty of Exact Sciences, \\ Tel Aviv University, Tel Aviv 69978,
Israel\\ }

\date{\today}
\maketitle
\begin{abstract}
Adiabatically pumped charge, carried by non-interacting electrons
through a quantum dot in a turnstile geometry, is studied as
function of the strength of the two modulating potentials (related
to the conductances of the two point-contacts to the leads) and of
the phase shift between them. It is shown that the magnitude and
sign of the pumped charge are determined by the relative position
and orientation of the closed contour traversed by the system in
the parameter plane, and the transmission peaks (or resonances) in
that plane. Integer values (in units of the electronic charge $e$)
of the pumped charge (per modulation period) are achieved when a
transmission peak falls inside the pumping contour. The integer
value is given by the winding number of the pumping contour:
double winding in the same direction gives a charge of 2, while
winding around two opposite branches of the transmission peaks or
winding in opposite directions can give a charge close to zero.

%Experimental determination of the resonance lines is suggested.

\end{abstract}


\begin{multicols}{2}

\section{Introduction and summary}

Quantized adiabatic charge transport was first invoked by
Thouless, \cite{Th83} who considered the dc current induced in an
infinite one-dimensional gas of non-interacting electrons by a
periodic potential, which is also a slowly-varying, periodic
function of time. The robustness of the quantization in such
systems, with respect to the influence of disorder or many-body
interactions, was further discussed in Refs. \onlinecite{NT84} and
\onlinecite{N90}. Since then, the focus of interest in this
phenomenon, termed ``electronic pump" \cite{He91,AG} has been
shifted to investigations of confined nanostructures, e.g.,
quantum dots or carbon nanotubes, where the realization of the
periodic time-dependent potential is achieved by modulating gate
voltages applied to the structure, in a ``turnstile" form,
\cite{Geerligs,Kou91,Po92,Sw99} or by coupling to surface acoustic
waves. \cite{Sh96,Ta97,Tal01} The pumping of charge in such
systems is achieved solely by the application of the
time-dependent potential, and it exists even when the system is
un-biased otherwise. Moreover, the charge transferred during a
single cycle is independent of the modulation period. However,
this charge is {\it not necessarily quantized}.

The quantization of the pumped charge, that is, the possibility to
transmit an integral number of electrons per cycle through an
un-biased system, has been realized in devices which are weakly
coupled to the leads, in which the (small) conductances of the
quantum point contacts separating the dot from the leads are
modulated in time. In such systems, the quantization has been
attributed to the Coulomb blockade, which quantizes the number of
electrons on the device. However, the modulation of the point
contact conductances allows, in principle, for the possibility to
cross-over from pinched-off conditions, where the Coulomb blockade
is effective, to the almost open-dot conditions. In the latter,
Coulomb-blockade effects are expected to play a minor role as
opposed to those of {\it quantum interference of the electronic
wave function} over the entire structure. Nonetheless, the
possible quantization of the charge pumped during a cycle through
a large, almost open quantum dot, with vanishing level spacing,
has been related \cite{AA98,SAA00} to Coulomb interactions within
the dot. Pumping in an open nanostructure was realized
experimentally for a quantum dot whose shape has been controlled
by oscillating gate voltages. \cite{Sw99} There, the amplitude of
the pumped signal, which is independent of the modulation
frequency, is found to increase with the driving force, though no
quantization has been detected.

While electronic correlations in fabricated nano-devices
definitely play a role, it is still of fundamental interest to
study electronic pumping of {\it non-interacting electrons}
resulting from quantum interference effects, to explore the
circumstances under which it is optimal. \cite{avron,makhlin} In
this context, it is especially useful to investigate simple,
tractable models, where it is possible to relate the parameters
characterizing the nanostructure, notably the conductance, with
those that govern the magnitude of the pumped charge. Thus by
considering a small, strongly pinched quantum dot, which supports
resonant transmission, it has been shown  \cite{Lev00} that when
the Fermi energy in the leads aligns with the energy of the
quasi-bound state in the dot, the charge pumped through it during
each period of the modulation is close to a single electronic
charge. The correlation between resonant states and enhanced
pumping has also been found in Ref. \onlinecite{W}. Finally, it
was pointed out \cite{W1} that the pumped charge in a model
describing carbon nanotubes can change sign as the amplitude of
the modulating potential is varied.

\begin{figure}
\leavevmode \epsfclipon \epsfxsize=7truecm
\vbox{\epsfbox{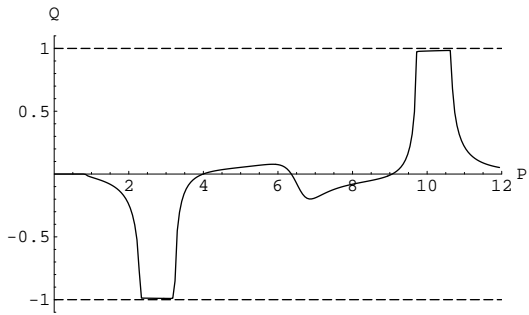}} \vspace{1cm} \caption{The pumped charge,
Q, in units of $e$, as function of the modulating amplitude, P,
for $N=4,~ka=0.001\pi,~J_D=J,~J_L=2 J,~\epsilon_0=0$ and
$\phi=0.4\pi$ (See text).}\label{fig1}
\end{figure}

In this paper we explore in detail the relation between resonant
transmission and {\it quantized} charge pumping. The results we
obtain can be summarized generically as follows. Consider a
quantum dot, connected by two single-channel point contacts to its
external leads. The individual conductances of these point
contacts, ${\rm X}_\ell$ and ${\rm X}_r$,  are controlled by
split-gate voltages which are modulated periodically in time.
Hence, during each cycle the system follows a closed curve in the
${\rm X}_\ell-{\rm X}_r$ parameter plane, which we call the
``pumping contour". As the various parameters (e.g. the modulation
amplitude P and phase shift $\phi$, see below) are varied, this
pumping contour changes its shape and location in the parameter
plane, forming a Lissajous curve. (This can be also achieved by
varying the gate voltage on the dot.) On the other hand, there are
lines in that parameter plane, along which the transmission of the
quantum dot is large. These are the ``resonance lines" of the
quantum dot. We find that the magnitude of the pumped charge and
{\it its sign} are intimately related to the manner by which the
pumping contour encircles parts of the resonance lines, and
particularly the peaks where the transmission is equal to unity.
The charge is quantized when a significant part of the resonance
line is trapped within the pumping contour. For example, when the
contour goes around the resonance twice, in the {\it same}
direction, the charge attains the value of $|2|$ in units of the
electronic charge $e$. The sign is determined by the sense of the
pumping contour. Thus, as function of the modulating amplitude,
the pumped charge can vary as depicted, for example, in Fig.
\ref{fig1}, or in Fig. \ref{fig7} bellow.

%\vspace{1cm}

The resonance lines in the parameter plane can be found
experimentally. \cite{Lev00}
%Let us denote the conductances of the
%point contacts by $G_{1}$ and $G_{2}$.
When one of the point contacts is pinched off and the other is
open, the conductance of the quantum dot, $G$, is dominated by the
conductance of the pinched-off point contact, say X$_\ell$.
Measuring the dependence of the quantum dot conductance on the
gate voltage $U_{\ell}$ controlling that point contact will yield
the relation between X$_{\ell}$ and $U_{\ell}$. In a similar way,
one finds the relation between X$_{r}$ and $U_{r}$. Using these
relations and measuring $G$ as function of $U_{\ell}$ and $U_{r}$,
the resonance lines can be determined.

To derive the above result we consider a simple model for a
quantum dot, employing the tight-binding description. In this
model, the quantum dot is coupled to semi-infinite 1D leads by
matrix elements $-J_{\ell}$ and $-J_{r}$, which oscillate in time
with frequency $\omega$, such that the modulation amplitude is P
and the phase shift between the $J_{\ell}$ modulation and that of
$J_{r}$ is $2\phi$.  Then, the point contact conductances are
given (in dimensionless units) by X$_{\ell}\equiv J_{\ell}^{2}$
and X$_{r}\equiv J_{r}^{2}$. Explicitly,
\begin{eqnarray}
J_{\ell}&=&J_{\rm L}+{\rm P}\cos (\omega t +\phi ),
\ \ {\rm X}_{\ell}\equiv J_{\ell}^{2},\nonumber\\
J_{\rm r}&=&J_{\rm L}+{\rm P}\cos (\omega t -\phi ),\ \ {\rm
X}_{r}\equiv J_{r}^{2} .\label{jljr}
\end{eqnarray}
 We solve for the charge pumped through
the quantum dot using the adiabatic approximation,
\cite{Th83,avron} that is, assuming that the frequency $\omega$ is
smaller than any characteristic energy scale of the electrons.
Concomitantly, we determine the resonance lines in the
{X$_{\ell}-{\rm X}_{r}$} plane. We then show that the pumping is
quantized (see Fig. \ref{fig1}) for values of P for which the
pumping contour encloses a significant part of a resonance line.

In our model, the couplings of the quantum dot  to the leads are
modulated in time and hence $J_{\ell}$ and $J_{r}$ can attain both
negative and positive values. This reflects a modulation of the
potential shaping the dot: the tight-binding parameters $J_\ell$
and $J_{\rm r}$, which are derived as integrals over the site
``atomic" wave functions and the oscillating potential, can have
both signs. The extreme modulation arises when we set $J_{\rm
L}=0$, so that the hopping matrix elements which couple the ``dot"
to the ``leads" oscillate in the range $\{-$P,P$\}$. The
dimensionless conductances X$_\ell$ and X$_r$ are then modulated
in the range $\{$0,P$^{2}\}$. The corresponding pumping contour is
then a simple closed elliptic curve. As the magnitude of the
average hopping $J_{\rm L}$  increases, the shape of the pumping
contour becomes more complex, turning into a Lissajous curve which
folds on itself. These complex pumping curves yield a rather rich
behavior of the pumped charge as function of the modulation
amplitude P (see e.g. Figs. 1 and 7).

\section{Transmission and pumping through a quantum dot}

Our description of the quantum dot is a generalization of the
model by Ng and Lee. \cite{Ng88} Imagine the quantum dot to be
connected to the electronic reservoirs by two 1D chains of sites,
whose on-site energies are assumed to vanish, and whose
nearest-neighbor transfer amplitudes are denoted by $-J$. The
energy of an electron of wave vector $k$ moving on such a chain is
\begin{eqnarray}
E_{k}=-2J\cos ka,
\end{eqnarray}
where $a$ is the lattice constant. From now on we measure energies
in units of $J$. The confined nanostructure, that is, the quantum
dot, is modeled by a `bunch' of tight-binding sites connected
among themselves. For simplicity, we take the latter in the form
of a finite chain of $N$ sites, each having the on-site energy
$\epsilon_{0}$, with a nearest-neighbor transfer amplitude
$-J_{\rm D}$. The effects of on-site interactions may be
considered as part of $\epsilon_0$, within a Hartree
approximation. This finite chain is attached to the left-hand-side
lead at site 1, with the matrix element $-J_{\ell}$, and to the
right-hand-side lead at site $N$, with the matrix element $-J_{r}$
[see Eq. (\ref{jljr})].

The calculation of the pumped charge requires the knowledge of the
instantaneous (that is, when the time is frozen) scattering
solutions of the problem, and the instantaneous scattering matrix.
\cite{B,we} This scattering matrix also yields the instantaneous
transmission coefficient of the quantum dot for any pair of the
parameters X$_{\ell}$ and X$_{r}$, and hence the resonance lines
in the X$_{\ell}-$X$_{r}$ plane. In the present case, the
instantaneous scattering solutions can be straightforwardly
obtained. Let us denote by $\chi^{t}_{\ell}$ the instantaneous
scattering state at time $t$, which is excited by an incoming wave
from the left reservoir, of energy $E_{k}$, and similarly
$\chi^{t}_{r}$ is the scattering solution excited by a wave coming
from the right. We then have
\begin{eqnarray}
\chi^{t}_{\ell}(x)&=&A_{0,\ell}\Bigl [e^{ikx}+r_{t}e^{-ikx}\Bigr ]
,\ {\rm on}\ {\rm the}\ {\rm
left}\ {\rm lead,}\nonumber\\
\chi^{t}_{\ell}(x)&=&A_{0,\ell}t_{t}e^{ikx},\ {\rm on}\ {\rm the}\
{\rm right}\ {\rm lead,}\label{xl}
\end{eqnarray}
where $x=na$, and similarly,
\begin{eqnarray}
\chi^{t}_{r}(x)&=&A_{0,r}\Bigl [e^{-ikx} +r'_{t}e^{ikx}\Bigr ],\
{\rm on}\ {\rm the}\ {\rm right}\ {\rm lead,}\nonumber\\
\chi^{t}_{r}(x)&=&A_{0,r}t'_{t}e^{-ikx},\ {\rm on}\ {\rm the}\
{\rm left}\ {\rm lead.}\label{xr}
\end{eqnarray}
To normalize the incoming waves such that they will carry a unit
flux, we put
\begin{eqnarray}
A_{0,\ell}=A_{0,r}=\sqrt{\frac{1}{2J\sin ka}}.
\end{eqnarray}
In Eqs. (\ref{xl}) and (\ref{xr}), $t_{t}=t'_t$, $r_{t}$ and
$r'_{t}$ are the instantaneous transmission and reflection
amplitudes, respectively. In our model, those are given by
\begin{eqnarray}
t_{t}&=&t'_t=-e^{-ik(N-1)a}J_{\ell}J_{r}J_{\rm D}\sin qa
M_{k},\nonumber\\
r_{t}&=&e^{i2ka}\Bigl [-1\nonumber\\
&+&\Bigl (e^{ika}{\rm X}_{\ell}{\rm X}_{r}\sin (N-1)qa -J_{\rm
D}{\rm X}_{\ell}\sin Nqa\Bigr
)M_{k}\Bigr ],\nonumber\\
r'_{t}&=&e^{-i2Nka}\Bigl [-1\nonumber\\
&+&\Bigl (e^{ika}{\rm X}_{\ell}{\rm X}_{r}\sin (N-1)qa -J_{\rm
D}{\rm X}_{r}\sin Nqa\Bigr )M_{k}\Bigr ].\label{scattering}
\end{eqnarray}
Here we have introduced the notations
\begin{eqnarray}
M_{k}&=&\frac{2i\sin ka}{D_{k}},\nonumber\\
D_{k}&=& J^{2}_{\rm D}\sin (N+1)qa
-J_{\rm D}e^{ika}({\rm X}_{\ell}+{\rm X}_{r})\sin Nqa\nonumber\\
&+&e^{i2ka}{\rm X}_{\ell}{\rm X}_{r}\sin (N-1)qa,
\end{eqnarray}
and scaled all energies in units of $J$. The wavevector $q$
describes the propagation of the wave on the quantum dot, such
that
\begin{eqnarray}
E_{k}-\epsilon_{0}=-2J_{\rm D}\cos qa .
\end{eqnarray}

\begin{figure}
\leavevmode \epsfclipon \epsfxsize=6truecm
\vbox{\epsfbox{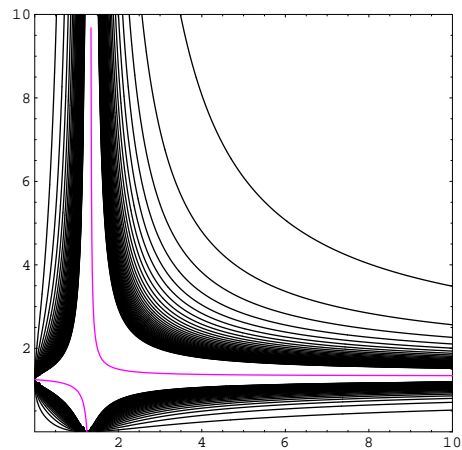}} \vspace{1cm} \caption{A contour plot of
the transmission in the X$_{\ell}-$X$_{r}$ plane, for the same
parameters as in Fig. \ref{fig1}. The resonance lines are the
colored curves in the figure. }\label{fig2}
\end{figure}

We begin the analysis by determining the resonance lines of the
transmission for our model. From the result for the transmission
amplitude $t_{t}$ [see Eqs. (\ref{scattering})], it is readily
found that the transmission coefficient is given by
\begin{eqnarray}
&&{\rm T}=|t_{t}|^{2}\nonumber\\
&=&\Bigl [1+\frac{{\rm Z}^{2}+( J_{\rm D}\sin ka\sin Nqa({\rm
X}_{\ell}-{\rm X}_{r}))^{2}}{(2J_{\rm D}\sin ka\sin qa)^{2}{\rm
X}_{\ell}{\rm X}_{r}}\Bigr ]^{-1},
\end{eqnarray}
with
\begin{eqnarray}
{\rm Z}&=&J_{\rm D}^{2}\sin (N+1)qa +\frac{E_{k}}{2}J_{\rm
D}\sin Nqa ({\rm X}_{\ell}+{\rm X}_{r})\nonumber\\
&+&{\rm X}_{\ell}{\rm X}_{r}\sin (N-1)qa .
\end{eqnarray}
Clearly, one has T=1 when X$_\ell={\rm X}_r$ and Z=0. For $N>1$,
these equations give two points on the diagonal in the
X$_\ell-$X$_r$ plane. It takes some algebra to show that the
second equation, Z$=0$, corresponds to maxima of T when either
X$_\ell$ or X$_r$ is varied while the other parameter is kept
fixed. These local maxima occur on two ``resonance lines", shown
in Fig. \ref{fig2}. This figure also shows a contour plot of the
transmission in the X$_{\ell}-$X$_{r}$ plane, for the same
parameters as in Fig. \ref{fig1}. For the chosen set of
parameters, the transmission is quite flat around the maximum; The
resonance lines appear as a ``ridge" on that plateau, with T
decreasing slowly as one moves away from the maximum points on the
diagonal along the resonance lines, but quickly as one moves away
from these lines.

Next we calculate the charge, Q, pumped through the quantum dot
during a single period of the modulation. Here we employ an
expression which gives Q in terms of matrix elements of the
temporal derivative of the scattering potential, $\dot{V}$,
between the instantaneous scattering states. \cite{we,aa} At zero
temperature the charge per period reads
\begin{eqnarray}
{\rm Q}=\frac{e}{4\pi}\oint dt\Bigl
[\langle\chi^{t}_{r}|\dot{V}|\chi^{t}_{r}\rangle
-\langle\chi^{t}_{\ell}|\dot{V}|\chi^{t}_{\ell}\rangle\Bigr ].
\label{charge}
\end{eqnarray}
This expression can be shown to reproduce the Brouwer \cite{B}
formula. \cite{we}

In our model, the temporal derivative of the potential is
\begin{eqnarray}
\dot{V}(n,n')&=&-\dot{J}_{\ell}\Bigl (\delta_{n,1}\delta_{n',0}
+\delta_{n',1}\delta_{n,0}\Bigr )\nonumber\\
&-&\dot{J}_{r}\Bigl
(\delta_{n,N}\delta_{n',N+1}+\delta_{n',N}\delta_{n,N+1}\Bigr ),
\end{eqnarray}
and hence the charge pumped from left to right is
\begin{eqnarray}
{\rm Q}&=&-\Re\Bigl \{\frac{e}{2\pi}\oint dt \Bigl (
\dot{J}_{\ell}\Bigl [\chi^{t\ast}_{r}(0)\chi^{t}_{r} (a)-
\chi^{t\ast}_{\ell}(0)\chi^{t}_{\ell} (a) \Bigr ]\nonumber\\
&+&\dot{J}_{r}\Bigl [
\chi^{t\ast}_{r}(Na)\chi^{t}_{r}((N+1)a)\nonumber\\
&-& \chi^{t\ast}_{ \ell}(Na)\chi^{t}_{\ell}((N+1)a)\Bigr ]\Bigr )
\Bigr \}.\label{q3}
\end{eqnarray}
This expression requires the knowledge of the scattering solutions
on the two ends of the quantum dot. These are given by using Eqs.
(\ref{xl}) and (\ref{xr}),
\begin{eqnarray}
\chi^{t}_{\ell}(a)&=&\frac{1}{J_{\ell}}A_{0,\ell}\Bigl
(e^{ika}+r_{t}e^{-ika}\Bigr ),\nonumber\\
\chi^{t}_{\ell}(Na)&=&\frac{1}{J_{r}}A_{0,\ell}t_{t}e^{ikNa},\nonumber\\
\chi^{t}_{r}(a)&=&\frac{1}{J_{\ell}}A_{0,r}t_{t}e^{-ika},\nonumber\\
\chi^{t}_{r}(Na)&=&\frac{1}{J_{r}}A_{0,r}\Bigl
(e^{-ikNa}+r'_{t}e^{ikNa}\Bigr ),
\end{eqnarray}
together with the results in Eq. (\ref{scattering}). It then
follows that the pumped charge, Q, is given by
\begin{eqnarray}
{\rm Q}&=&\frac{e}{2\pi}\oint dt\frac{J_{\rm D}\sin Nqa\sin
ka}{|D_{k}|^{2}}\nonumber\\
&&\times\Bigl [\Bigl ({\rm X}_{\ell}\dot{\rm X}_{r}-{\rm
X}_{r}\dot{\rm X}_{\ell}\Bigr )J_{\rm D}E_{k}\sin
Nqa\nonumber\\
&+&\Bigl ({\rm X}^{2}_{\ell}\dot{\rm X}_{r}-{\rm
X}_{r}^{2}\dot{\rm X}_{\ell}\Bigr )\sin (N-1)qa\nonumber\\
&+&\Bigl (\dot{\rm X}_{r}-\dot{\rm X}_{\ell}\Bigr )J_{\rm
D}^{2}\sin (N+1)qa\Bigr ].\label{Q}
\end{eqnarray}
The temporal integration in this expression, with the time
dependence as given by Eq. (\ref{jljr}), is now done analytically,
separating the four poles of the denominator in $\cos \omega t$.
The results, for a selected set of parameters, are shown in Fig.
\ref{fig1}.
%The parameters used in the plot are
%$\epsilon_{0}=0$, $J_{\rm D}=1$, $J_{\rm L}=2$, $ka=0.001\pi$,
%$\phi =0.4\pi$, and $N=4$.

\begin{figure}
\leavevmode \epsfclipon \epsfxsize=7truecm
\vbox{\epsfbox{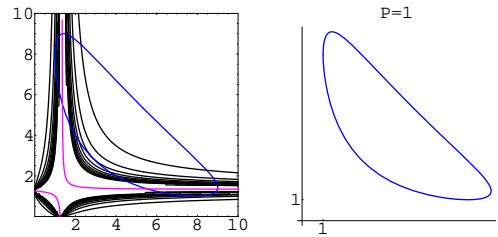}} \vspace{1cm} \caption{The pumping curve
in the X$_{\ell}-$X$_{r}$ plane, for P=1. The other parameters are
the same as in Fig. \ref{fig1}. The right plate shows the full
pumping contour; The left one depicts the contours of equal
transmission (in black), the resonance lines (in red), and the
pumping contour (in blue). }\label{fig3}
\end{figure}

\begin{figure}
\leavevmode \epsfclipon \epsfxsize=7truecm
\vbox{\epsfbox{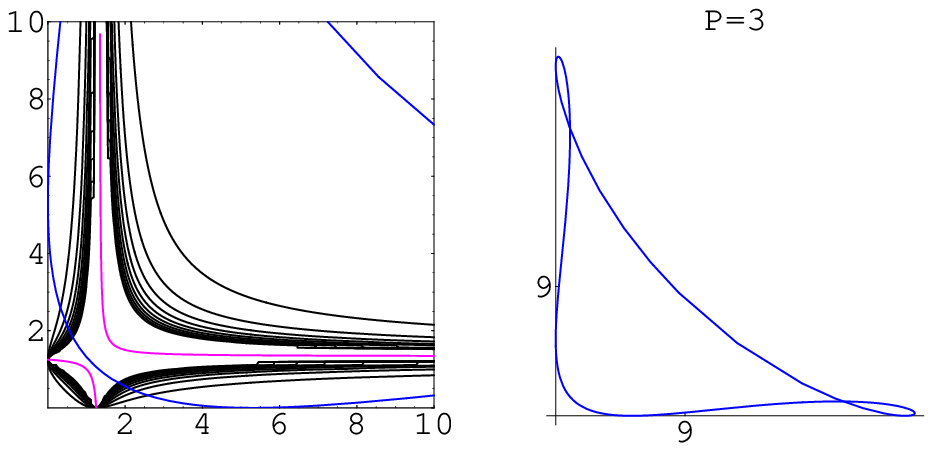}} \vspace{1cm} \caption{The same as Fig.
\ref{fig3}, but for P=3.  The other parameters are the same as in
Fig. \ref{fig1}.}\label{fig4}
\end{figure}

\begin{figure}
\leavevmode \epsfclipon \epsfxsize=7truecm
\vbox{\epsfbox{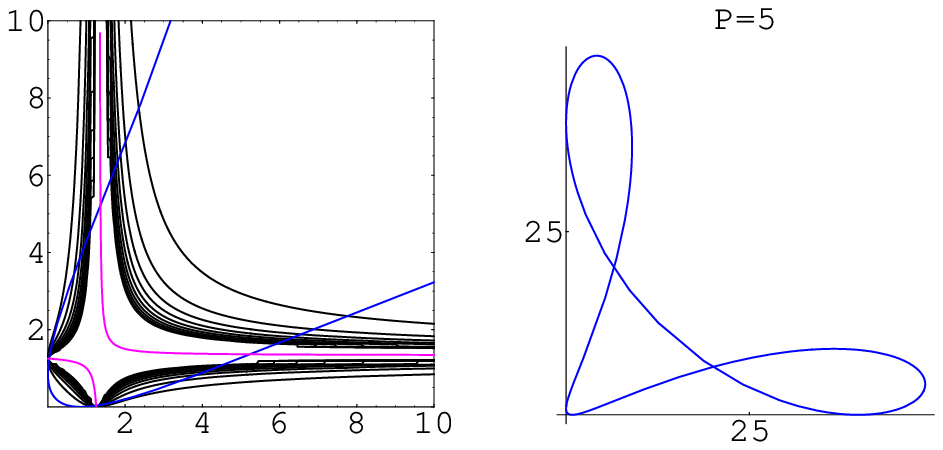}} \vspace{1cm} \caption{The same as Fig.
\ref{fig3}, but for P=5. The other parameters are the same as in
Fig. \ref{fig1}.}\label{fig5}
\end{figure}

\begin{figure}
\leavevmode \epsfclipon \epsfxsize=7truecm
\vbox{\epsfbox{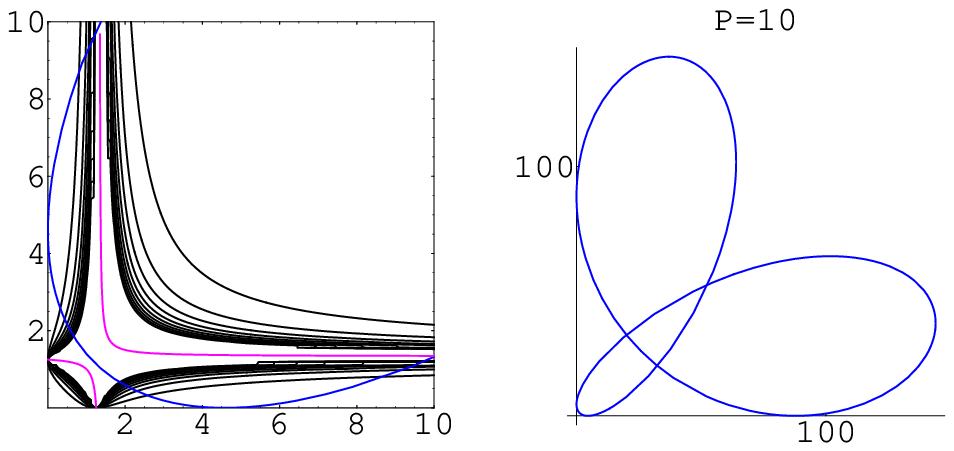}} \vspace{1cm} \caption{The same as Fig.
\ref{fig3}, but for P=10. The other parameters are the same as in
Fig. \ref{fig1}.}\label{fig6}
\end{figure}

We now relate the values of Q, for representative values of the
pumping amplitude P, to the pumping contour, which is the closed
curve the system traverses during a single cycle in the
X$_{\ell}-$X$_{r}$ plane. Those curves are portrayed in Figs.
\ref{fig3} (P=1), \ref{fig4} (P=3), \ref{fig5} (P=5), and
\ref{fig6} (P=10). In each of the these figures, the left plate
shows the contours of equal transmission (in black), the resonance
lines (in red) and that section of the pumping contour which lies
near the transmission maxima (in blue). The right plate exhibits
the full pumping contour. (The numbers on the axes in the right
plate indicate the scale of the pumping contour.) We begin with
the situation at P=1. At this value, as can be seen from Fig.
\ref{fig1}, the charge pumped is vanishingly small. At this value,
the pumping contour indeed encloses only a small portion of the
resonance line, far away from the transmission maxima (see Fig.
\ref{fig3}).
%\noindent

Next, we consider the case P=3, for which the pumped charge is
close to $-1$ (in units of $e$). Examining Fig. \ref{fig4}, it is
seen that in this case the pumping contour encloses  only the
upper resonance line, including the transmission peak there. For
this value of P, the pumping contour begins to fold on itself
(since X$_{\ell}$ and X$_{r}$ are definitely positive), forming a
Lissajous curve in that parameter plane. This tendency is of
course enhanced as the pumping amplitude P increases. The charge
remains close to $-1$ as long as the pumping contour remains
between the two branches of the resonance line, as in Fig.
\ref{fig4}.
Moving to the case P=5 (Fig. \ref{fig5}), at which the pumped
charge is small and positive (see Fig. \ref{fig1}), we find that
now the pumping contour encircles {\it both} transmission peaks on
the resonance lines; The separate contributions of the two
resonances to the temporal integration in Eq. (\ref{Q}) almost
cancel one another, leading to a rather tiny value for the charge.
Finally, we consider the case P=10. The left plate in Fig.
\ref{fig6} is similar to the left one in Fig. \ref{fig4}, and
indeed, the absolute value of Q is close to the one in both cases.
However, the sense in which the pumping contour encircles the
resonance is reversed in the two cases, as can be gathered by
examining the right plates in Figs. \ref{fig4} and \ref{fig6}.
Hence, the pumped charge changes sign.

\begin{figure}
\leavevmode \epsfclipon \epsfxsize=7truecm
\vbox{\epsfbox{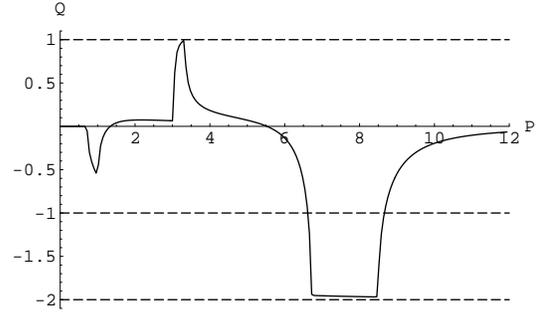}} \vspace{1cm} \caption{The pumped charge,
Q, in units of $e$, as function of the modulating amplitude, P.
The parameters are the same as in Fig. \ref{fig1}, except that
$\phi=0.05 \pi$. }\label{fig7}
\end{figure}

\begin{figure}
\leavevmode \epsfclipon \epsfxsize=7truecm
\vbox{\epsfbox{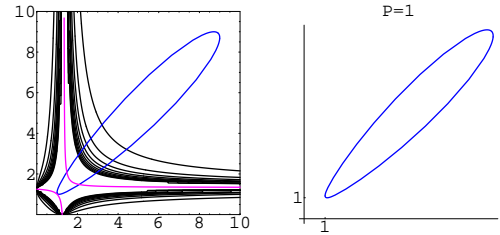}} \vspace{1cm} \caption{The pumping
curve in the X$_{\ell}-$X$_{r}$ plane, for P=1. The other
parameters are the same as in Fig. \ref{fig7}. The right plate
shows the full pumping curve; The left one depicts the contours of
equal transmission (in black), the resonance lines (in red), and
the pumping contour (in blue).  }\label{fig8}
\end{figure}

\begin{figure}
\leavevmode \epsfclipon \epsfxsize=7truecm
\vbox{\epsfbox{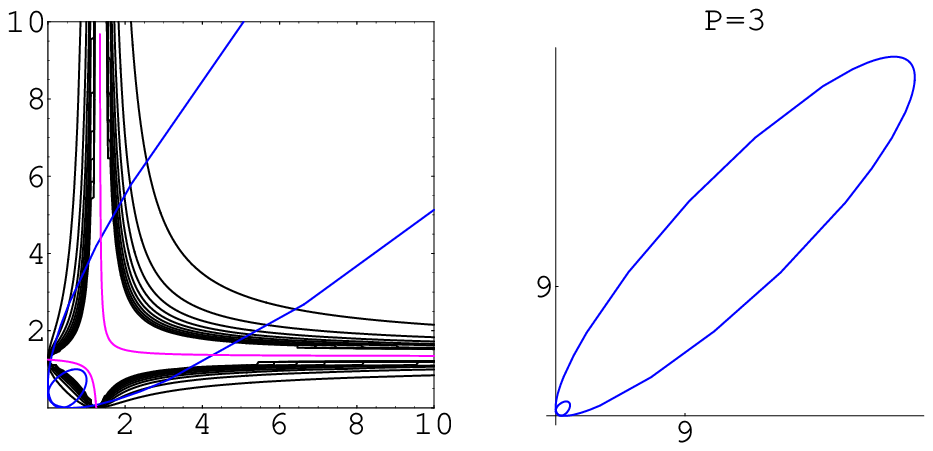}} \vspace{1cm} \caption{The same as
Fig. \ref{fig8}, but for P=3. The other parameters are the same as
in Fig. \ref{fig7}. }\label{fig9}
\end{figure}

\begin{figure}
\leavevmode \epsfclipon \epsfxsize=7truecm
\vbox{\epsfbox{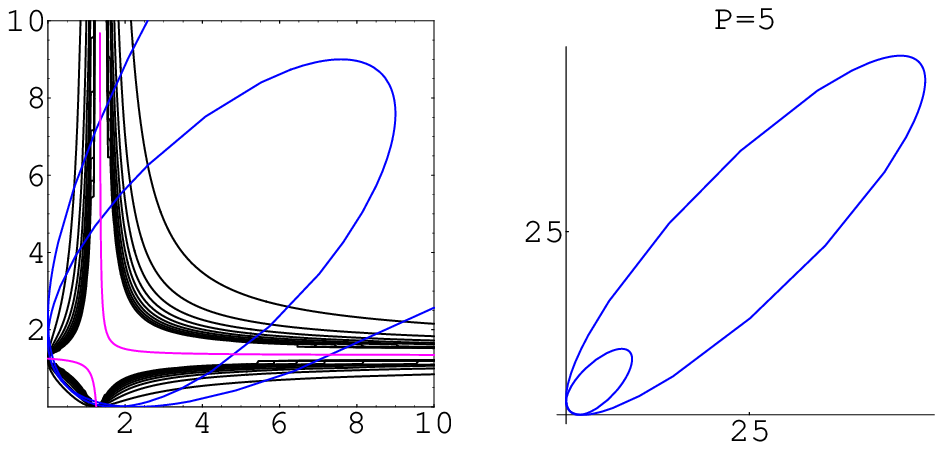}} \vspace{1cm} \caption{ The same as
Fig. \ref{fig8}, but for P=5. The other parameters are the same as
in Fig. \ref{fig7}.}\label{fig10}
\end{figure}

\begin{figure}
\leavevmode \epsfclipon \epsfxsize=7truecm
\vbox{\epsfbox{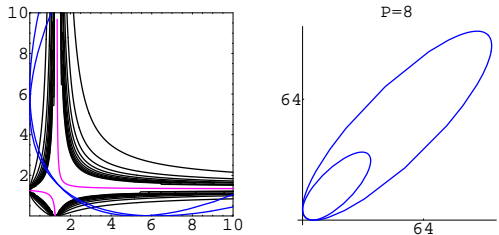}} \vspace{1cm} \caption{ The same as
Fig. \ref{fig8}, but for P=8. The other parameters are the same as
in Fig. \ref{fig7}.}\label{fig11}
\end{figure}

\begin{figure}
\leavevmode \epsfclipon \epsfxsize=7truecm
\vbox{\epsfbox{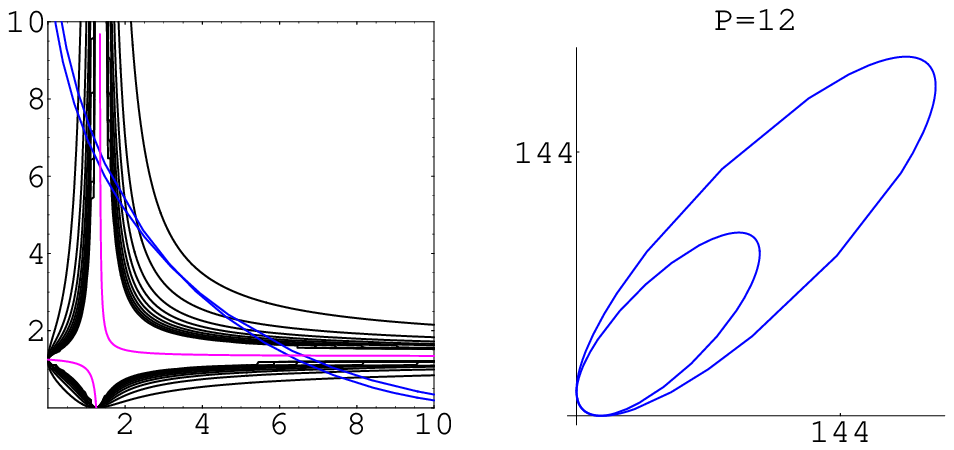}} \vspace{1cm} \caption{The same as
Fig. \ref{fig8}, but for P=12. The other parameters are the same
as in Fig. \ref{fig7}. }\label{fig12}
\end{figure}

We now examine the pumping when the phase shift $\phi$ of the
modulation [see Eq. (\ref{jljr})] is changed. Instead of taking
$\phi =0.4 \pi$, which was used to plot the previous figures, we
now choose $\phi =0.05 \pi$, keeping all other parameters as
before. Although one might have thought that this smaller value
will reduce the magnitude of the pumped charge, we now find that
$|{\rm Q}|$ attains the value of $2$, see Fig. \ref{fig7}.
The next five figures portray the pumping contours relation to the
resonance lines for representative values of the amplitude P, for
the parameters of Fig. \ref{fig7}. In Fig. \ref{fig8} we have P=1.
Then the pumping contour encloses just a small part of the upper
resonance line, and also touches the peak on the lower resonance
line. Indeed, Q has an intermediate value near $-0.5$, decreasing
to zero as P moves away from 1 (see Fig. \ref{fig7}).
Increasing the amplitude to the value P=3 reveals that the pumping
contour encloses both peaks on the resonance lines, and therefore
their separate contributions almost cancel one another, leading to
a tiny value of Q. Also, the pumping curve begins to fold on
itself, giving rise to the small ``bubble" close to the origin
(see Fig. \ref{fig9}). Following the increase of that bubble as P
in enhanced leads to the situation in which the bubble encloses
the lower resonance, and then the charge attains the value 1 (see
Fig. \ref{fig7}). As the bubble increases further, capturing the
two resonance lines, as shown in Fig. \ref{fig10}, Q again becomes
very small.
As P grows on, we reach the interesting situation, depicted in
Fig. \ref{fig11} for P=8, in which the bubble encircles {\it
twice} the upper resonance line, leading to a pumped charge very
close to $|2e|$.
Finally, Fig. \ref{fig12} shows the pumping contour for P=12. The
contour is seen to shift away from the resonance peaks, yielding a
vanishingly small value for Q.

\section{Conclusions}

We have calculated the charge passing through a small quantum dot
whose point contact conductances are modulated periodically in
time. The calculation is carried out for non-interacting
electrons, under the assumption that the modulation frequency is
much smaller than the electronic relaxation rates. We have found
the conditions for obtaining integral values for the pumped
charge: The contour traversed by the system in the parameter plane
spanned by the pumping parameters (in our case, the point contact
conductances) should encircle a significant portion of a resonance
line (along which the transmission of the quantum contact is
optimal) in that plane. The magnitude and the sign of the pumped
charge are determined by that portion, and by the direction along
which the resonance line is encompassed.

The reason for this observation can be traced back to the
expression for the pumped charge, Eqs. (\ref{charge}) and
(\ref{Q}). The contribution to the temporal integration comes
mainly from the poles in the denominator in Eq. (\ref{Q}). These
same poles are also responsible for the resonant states of the
nanostructure, that is, for the maxima in the transmission
coefficient. \cite{Lev00,we,aa}   In this way, we have obtained a
topological description for the phenomenon of adiabatic charge
pumping. One can now imagine more complex scenarios: including
higher harmonics of $\omega$ in the time dependence of the point
contact conductances can create more complex Lissajoux  contours,
which might encircle portions of the resonance lines more times,
yielding higher quantized values of the pumped charge.

 Finally, it
should be mentioned that the results presented above are obtained
at zero temperature. At finite temperatures, the expression for Q
should be integrated over the the electron energy $E_{k}$, with
the Fermi function derivative $-\partial f/\partial E_
{k}$.\cite{we,aa} Hence, upon the increase of the temperature, the
pumped charge would be smeared and suppressed. It would be very
interesting to check the above predictions in more complicated
models (for example, when there are several levels on each of the
sites forming the quantum dot), allowing for a richer structure of
the transmission in the parameter space, and, of course, in real
systems.

\acknowledgements

This research was carried out in a center of excellence supported
by the Israel Science Foundation, and was supported in part by the
National Science Foundation under Grant No. PHY99-07949, and by
the Albert Einstein Minerva Center for Theoretical Physics at the
Weizmann Institute of Science.

\end{multicols}

\end{document}